\documentclass[10pt, conference, compsocconf]{IEEEtran}
\IEEEoverridecommandlockouts
\usepackage{cite}
\usepackage{amsmath,amssymb,amsfonts}
\usepackage{algorithmic}
\usepackage{graphicx}
\usepackage{textcomp}
\usepackage{xcolor}
\def\BibTeX{{\rm B\kern-.05em{\sc i\kern-.025em b}\kern-.08em
    T\kern-.1667em\lower.7ex\hbox{E}\kern-.125emX}}

\usepackage{times}
\usepackage{xcolor}
\usepackage{soul}
\usepackage[utf8]{inputenc}
\usepackage[small]{caption}
\usepackage{graphicx}
\usepackage{listings}

\usepackage{amsmath}
\usepackage{amsfonts}
\usepackage{amssymb}
\usepackage{amsthm}

\usepackage{amscd}
\usepackage[utf8]{inputenc}
\usepackage{stmaryrd}

\newtheorem{defn}{Definition}
\newtheorem{prop}[defn]{Proposition}

\newtheorem{corolary}[defn]{Corollary}

\newcommand{\ALG}{\ensuremath{\mathbf{Algorithm}}}
\newcommand{\INPUT}{\ensuremath{\mathbf{Input:}}}
\newcommand{\OUTPUT}{\ensuremath{\mathbf{Output:}}}

\newcommand{\IF}{\ensuremath{\mathbf{if}}}

\newcommand{\THEN}{\ensuremath{\mathbf{then}}}
\newcommand{\ELSE}{\ensuremath{\mathbf{else}}}

\newcommand{\RETURN}{\ensuremath{\mathbf{return}}}
\newcommand{\FOR}{\ensuremath{\mathbf{for}}}

\newcommand{\FOREACH}{\mbox{\textbf{for each}}}

\newcommand{\AND}{\ensuremath{\mathbf{and}}}
\newcommand{\LIN}[1]{\textbf{[#1]}}

\renewcommand{\citeonline}[1]{\cite{#1}}

\begin{document}
\title{Tractable reasoning about Agent Programming in Dynamic Preference Logic}

\author{\IEEEauthorblockN{Marlo Souza}
\IEEEauthorblockA{Institute of Mathematics and Statistics\\
UFBA\\
Salvador, Brazil\\
Email: marlo@dcc.ufba.br}
\and
\IEEEauthorblockN{Álvaro Moreira}
\IEEEauthorblockA{Institute of Informatics\\
UFRGS\\
Porto Alegre, Brazil\\
Email: afmoreira@inf.ufrgs.br}
\and
\IEEEauthorblockN{Renata Vieira}
\IEEEauthorblockA{Faculty of Informatics\\
PUCRS\\
Porto Alegre, Brazil\\
Email: renata.vieira@pucrs.br}
}

\maketitle

\begin{abstract}
While several BDI logics have been proposed in the area of Agent Programming, it is not clear how these logics are connected to the agent programs they are supposed to specify. More yet, the reasoning problems in these logics, being based on modal logic, are not tractable in general, limiting their usage to tackle real-world problems. In this work, we use of Dynamic Preference Logic to provide a semantic foundation to BDI agent programming languages and investigate tractable expressive fragments of this logic to reason about agent programs. With that, we aim to provide a way of implementing semantically grounded agent programming languages with tractable reasoning cycles.
\end{abstract}

\section{Introduction}

In the study of rational action and agency, several different logics and formal theories for practical reasoning have been proposed. Particularly, the Belief, Desire Intention framework \cite{bratman} has become a popular approach to practical reasoning in the areas of Artificial Intelligence and Autonomous Agents, giving rise to the construction of several programming languages and computer systems.

Having a formal definition of its semantics is essential for proving properties about a programming language and also for providing a formal framework to specify and to verify properties about a system's behaviour. For agent programming languages, a formal semantics also clarifies the notion of agency it carries.

Recently, it has been proposed that Dynamic Preference Logic (DPL) can be used to reason about BDI Agent Programming with declarative mental attitudes \cite{souza2017dynamic}, providing a. 
a computable two-way translation between specifications in the logic and agent programs. Reasoning in DPL is, however, not tractable in general.
As such, the use of DPL for reasoning about agent programming - while theoretically relevant for the analysis of a programming language semantics - is of very limited practical use.

In this work, we investigate expressive fragments of the language of DPL that yield tractable reasoning problems. The reasoning problems discussed in this paper are concerned with knowing whether an agent knows (believes, desires or intends) a certain propositional property $\varphi$ and how to compute the resulting mental state of an agent after performing a belief/desire revision or contraction. With this, we aim to provide a tractable fragment that may be used to implement an actual agent programming language with declarative mental attitudes having a well-defined logical semantics based on Kripke frames. 

This work is structured as follows: in Section~\ref{sec:bdi} we present the logic of agency proposed here, based on Dynamic Preference Logic; in Section~\ref{sec:reas} we discuss the connection between the logic proposed and Agent Programming. In Section~\ref{sec:trac}, we discuss a tractable expressive subset of the language which can be used to implement agent programming languages. In section~\ref{sec:relwork}, we present the related work and compare their contributions to ours. Finally, in Section~\ref{sec:final}, we present our final considerations.

\section{A dynamic logic for BDI programming}\label{sec:bdi}

In this section, we present a dynamic propositional modal logic of agency which will be used to specify an agent's mental state. 
Throughout this work, we will assume the existence of a set $P$ of propositional symbols and we will denote by $\mathcal{L}_0$ the language of propositional logic constructed over the symbols in $P$. 

We assume a BDI agent has a library of plans describing which actions she can perform on the environment. For the sake of simplicity, we will assume that the plans are deterministic, completely specified and STRIPS-like. 
 Aware of these restrictions, we introduce the notion of plan library.

\begin{defn}\label{def:ra}
We call $\mathcal{A} = {\langle \Pi, pre, pos\rangle}$ 
a plan library, iff $\Pi$ is a finite set of plans symbols, $pre,pos: \Pi \rightarrow \mathcal{L}_0$ are functions that map each plan to its preconditions and post-conditions, respectively. 
We further require that  the post-conditions of any plan is a consistent conjunction of propositional literals. 
We say $\alpha \in \mathcal{A}$ for any plan symbol $\alpha \in \Pi$.
\end{defn}

With this definition in mind, we can establish the language we will use as a base for our constructions.

\begin{defn}\label{def:lang}
Let $\mathcal{A}={\langle \Pi, pre, pos\rangle}$ be a plan library. We define the language $\mathcal{L}_{\leq_P,\leq_D}( \mathcal{A})$ by the following grammar (where $p \in P$ and $\alpha \in \mathcal{A}$): $$
\begin{array}{lc}
\varphi ::=& p ~|~ \neg \varphi ~|~ \varphi \wedge \varphi ~|~ A \varphi ~ | ~ [\leq_P]\varphi ~|~ [<_P] \varphi~\\
&~| ~ [\leq_D]\varphi ~|~ [<_D] \varphi~|~ [\alpha]\varphi ~|~ I\alpha
\end{array}$$
\end{defn}

The formula $A\varphi$ means that it is universally true that $\varphi$ holds, while the formula $[\leq_D]\varphi$ (${[\leq_P]\varphi}$) means that in all worlds equally or more desirable (plausible) than the current one, $\varphi$ holds and $[<_D]\varphi$ ($[<_P]\varphi$) meaning that in all worlds strictly more desirable (plausible) than the current one, $\varphi$ holds. Finally, the formulas $[\alpha]\varphi$ and $I\alpha$ mean that after carrying out the plan $\alpha$, the property $\varphi$ holds, and that the agent intends to execute a plan $\alpha$, respectively. As usual, we will define $E\varphi \equiv \neg A \neg \varphi$ and ${\langle \leq_\Box \rangle} \varphi \equiv \neg {[\leq_\Box]}\neg\varphi$ with $\Box \in \{P,D\}$. 

To interpret these formulas, we introduce 
the notion of \textit{agent model} - which will be used to represent an agent's mental state.

\begin{defn}\label{def:agmodel}
Let $\mathcal{A} = \langle \Pi, pre, pos\rangle$ be a plan library. An agent model is a tuple $M = {{\langle W, \leq_P,\leq_D, I,v\rangle}}$ where $W$ is a set of possible worlds, and both $\leq_P$ and $\leq_D$ are pre-orders over $W$ with well-founded strict parts $<_P$ and $<_D$, $I \subseteq \Pi$ is a set of adopted plans (or intentions), and $v$ is a valuation function.
\end{defn} 

To model the effect of executing a plan $\alpha \in \mathcal{A}$ given an agent model $M$, we will define the notion of model update, as commonly used in the area of Dynamic Epistemic Logic. 

\begin{defn}
Let $\mathcal{A} = {\langle \Pi,  pre, pos \rangle}$ be a plan library, $\alpha \in \mathcal{A}$ a plan, and $M= {\langle W, \leq_P,\leq_D, I, v\rangle}$ an agent model. The update of model $M$ by execution of plan $\alpha$ is defined as the model $M \otimes [\mathcal{A},\alpha] = {\langle W', \leq_P', \leq_D', I', v'\rangle}$ where
$$\begin{array}{lll}
\small
W' & = & \{w\in W ~|~ M, w\vDash pre(\alpha)\}\\
\leq_\Box' & = & \leq_\Box \cap~ W'\times W'\\
I' & =& I\\
v'(p) &=& \begin{cases}
					W' & \mbox{if }pos(\alpha)\vDash p\\
					\emptyset &\mbox{if }pos(\alpha)\vDash \neg p\\
					v(p) \cap~ W'& \mbox{otherwise}
\end{cases}
\end{array}$$
\end{defn}

The interpretation of the formulas is defined as usual, with each modality corresponding to an accessibility relation. 
$$\begin{array}{ll}
M,w \vDash [\leq_P]\varphi & \mbox{ iff }\forall w' \in W: w'\leq_P w \Rightarrow M,w' \vDash  \varphi\\
M,w \vDash [<_P]\varphi & \mbox{ iff }\forall w' \in W: w'<_P w \Rightarrow M,w' \vDash \varphi\\
M,w \vDash [\leq_D]\varphi & \mbox{ iff }\forall w' \in W: w'\leq_D w \Rightarrow M,w' \vDash  \varphi\\
M,w \vDash [<_D]\varphi & \mbox{ iff }\forall w' \in W: w'<_D w \Rightarrow M,w' \vDash \varphi\\
M,w \vDash [\alpha]\varphi & \mbox{ if } M,w \vDash pre(\alpha) \mbox{ then } M\otimes [\mathcal{A},\alpha],w \vDash \varphi\\
M,w \vDash I\alpha &\mbox{ iff } \alpha \in I 
\end{array}$$

In this work, we denote the most plausible (similarly, desirable) worlds in the set $S$ by $Min_{\leq_P} S$. In other words, $$Min_{\leq_P} S= \{w\in S~|~ \forall w'\in S: w'\leq_P w \Rightarrow w\leq_P w'\}$$
 
Let $\varphi \in \mathcal{L}_{\leq_P,\leq_D}(\mathcal{A})$, we also define the formula $\mu_P \varphi~\equiv~(\varphi \wedge \neg \langle <_P \rangle \varphi)$ (similarly, $\mu_D \varphi$), which is satisfied exactly by the minimal worlds according to the order $\leq_P$ (similarly for $\leq_D$) which satisfy the formula $\varphi$, i,e, $\llbracket \mu_P \varphi \rrbracket = Min_\leq \llbracket \varphi\rrbracket$. These formulas will be useful to encode mental attitudes in this logic.

\subsection{Enconding mental attitudes}

Following Souza et al~\cite{souzaphd}\cite{souza2017dynamic}, we introduce a codification of mental attitudes in the language $\mathcal{L}_{\leq_P, \leq_D}( \mathcal{A})$. In this work, we interpret the notion of `possible world' as epistemically possible, not metaphysically possible. As such, the universal modality can be used to encode the knowledge held by an agent.
$$K\varphi \equiv  A\varphi$$

We encode the (KD45) notion of belief as what is true in the worlds that the agent believes to be the most plausible ones. As such, our notion $B\varphi$ means that `\textit{it is most plausible that} $\varphi$.'
\[
B\varphi \equiv A(\mu_P \top\rightarrow \varphi)\]


Similar to belief, we propose a codification of (consistent) desires as everything that is satisfied in all most desirable worlds. 
\[G(\varphi) \equiv A(\mu_D \top \rightarrow \varphi)\]

Our language possesses the notion of procedural intentions by the formula $I\alpha$. To encode Bratman's~\citeonline{bratman} notion of prospective intention, we will define a formula $Int\varphi$. First, however, we must encode the restrictions imposed by Bratman for consistency of an intention by means of a formula $AdmInt(\varphi)$ meaning that `\textit{it is admissible for the agent to intend that} $\varphi$'.

\[
AdmInt(\varphi)\equiv G(\varphi)\wedge E(\varphi)\wedge \neg B(\varphi)
\]

With this notion, we can define the notion of having an `\textit{intention that} $\varphi$.'
\[
Int (\varphi) \equiv AdmInt(\varphi)\wedge \bigvee_{\alpha \in \mathcal{A}} (I\alpha \wedge B\left(pre(\alpha)\wedge [\alpha]\varphi \right))
\]

Notice that, while we imposed several restrictions in our codification for an agent to rationally hold some prospective intention, none of these restrictions have been required for an agent to hold a procedural intention, i.e. an \textit{intention to do} - here represented by the set of adopted plans $I$ in the model. 
To model the kind of agent that satisfies Bratman's restrictions, we define the notion of a coherent agent model, i.e. an agent having a coherent mental state.

\begin{defn}
Let $\mathcal{A}$ be a plan library and $M = {\langle W, \leq_P,\leq_D, I ,v\rangle}$ be an agent model. We say a set $I\subset \Pi$ of plans is $\mathcal{A}$-coherent in $M$ if for all $\alpha \in I$, $M\vDash B(pre(\alpha))$ and $M\vDash AdmInt(pos(\alpha))$. If $I$ is $\mathcal{A}$-coherent in $M$, we say $M$ is a coherent agent model.
\end{defn}


\subsection{Dynamic operations on agent's mental states }

Once established the basic language and the encodings of mental attitudes, we define some well-behaved mental operations, which will be used to implement agents' practical reasoning. 
Here we explore three dynamic operations on agent models, each a representative of the three basic mental operations as studied by the Belief Revision Theory \cite{AGM85}: expansion, revision and contraction.

The first operation we introduce is that of public announcement. This operation corresponds, in a sense, to the operation of expansion from Belief Revision Theory. Based on the codifications we provided in the previous section, this operation can be interpreted as the mental operation of knowledge acquisition.

\begin{defn}\label{def:PA}
\sloppy
Let $M= {\langle W, \leq_P,\leq_D, I, v\rangle}$ be a coherent agent model and $\varphi$ a formula of $\mathcal{L}_0$. We say the model  $M_{!\varphi} = {\langle W_{!\varphi}, \leq_{P!\varphi}, \leq_{D!\varphi}, I_{!\varphi}, v_{!\varphi}\rangle}$ is the result of public announcement of $\varphi$ in $M$,  where: 

$
\begin{array}{lll}
W_{!\varphi} &=& \{w \in W ~|~ M,w\vDash \varphi\}\\
\leq_{\Box!\varphi} &=& \leq_\Box \cap ~(W_{!\varphi}\times W_{!\varphi})\\
I_{!\varphi} & \multicolumn{2}{l}{\mbox{is the maximal subset of } I \mbox{ that is }\mathcal{A}-\mbox{coherent}}\\
v_{!\varphi}(p) &=& v(p)~\cap~ W_{!\varphi} 
\end{array}
$%
\end{defn}

%

The radical upgrade of an agents beliefs by an information $\varphi$ results in a model such that all worlds satisfying $\varphi$ are deemed more plausible than those not satisfying it. This operation corresponds to a operation of belief revision from belief Revision Theory. 

\begin{defn}
\sloppy
\label{def:RUp}
Let $M= {\langle W, \leq_P,\leq_D, I, v\rangle}$ be a coherent agent model and $\varphi$ a formula of $\mathcal{L}_0$. We say the model  $M_{\Uparrow\varphi} = {{\langle W, \leq_{P\Uparrow\varphi},\leq_D,I_{\Uparrow_P \varphi},  v\rangle}}$ is the result of the radical upgrade on the plausibility of $M$ by $\varphi$,  where

{\centering $
\begin{array}{ll}
\leq_{P\Uparrow \varphi} = & (\leq_P \setminus \{{\langle w,w' \rangle}~|~ M,w\not\vDash\varphi \mbox{ and }M,w'\vDash\varphi \})\cup\\
&\{{\langle w,w' \rangle} ~|~ M,w\vDash\varphi \mbox{ and }M,w'\not\vDash\varphi \}\\
I_{\Uparrow_P \varphi}  & \mbox{is the maximal subset of } I \mbox{ that is }\mathcal{A}-\mbox{coherent}\\
\end{array}$}
\end{defn}

We can similarly define the radical upgrade of the agents desires by the operation $\Uparrow_D \varphi$, which updates the desirability relation, instead of the plausibility relation.

%

Lastly, we introduce the operation of lexicographic contraction \cite{ramachandran2012three}. This operation corresponds to performing a re-ordering of the worlds in a way that both $\varphi$ and $\neg \varphi$ are considered equally plausible (or equally desirable) to the agent. To define this operation in a more elegant way, we will define the notion of the plausibility degree of a world. 

\begin{defn}\label{def:deg}
\sloppy
Let $M= {\langle W, \leq_P,\leq_D, I, v\rangle}$ be a coherent agent model, $w\in W$ a possible world and $\varphi\in \mathcal{L}_{\leq_P , \leq_D}(\mathcal{A})$ a formula. We say that $w$ has plausibility degree $n \in \mathbb{N}$ for $\phi$, denoted $n = {d_P}_\varphi(w)$, if $M,w\vDash \varphi$ and there is a maximal chain 
$w_0 <_P w_1 <_P w_2 <_P\ldots <_P w_n$ s.t. $\forall i: w_i \in \llbracket \varphi\rrbracket$, $w_0 \in Min_{\leq_P} \llbracket \varphi\rrbracket$ and $w_n = w$.
\end{defn}

We can define the desirability degree of a world $w$ for $\varphi$, denoted by ${d_D}_\varphi(w)$, the same way. 
With that notion, we define the lexicographic contraction as below. 

\begin{defn}
\label{def:lexcont}
\sloppy
Let $M= {\langle W, \leq_P,\leq_D, I, v\rangle}$ be a coherent agent model and $\varphi$ a formula of $\mathcal{L}_0$. We say the model  $M_{\Downarrow_P\varphi} = {{\langle W, \leq_{P\Downarrow\varphi},\leq_D,I_{\Downarrow_P \varphi},  v\rangle}}$ is the lexicographic contraction of the plausibility of $M$ by $\varphi$,  where: 

${
\centering
\footnotesize
\begin{array}{l}
w\leq _{P\Downarrow\varphi} w'\mbox{ iff }\begin{cases} w \leq_P w' & \mbox{if }w,w' \in \llbracket \varphi \rrbracket\\
w \leq_P w' & \mbox{if }w,w' \not\in \llbracket \varphi \rrbracket\\
{d_P}_\varphi(w) < {d_P}_{\neg\varphi}(w') & \mbox{if }w \in \llbracket \varphi \rrbracket \mbox{ and }\\ &\quad w' \not\in \llbracket \varphi \rrbracket\\
{d_P}_{\neg \varphi}(w) < {d_P}_{\varphi}(w') & \mbox{if }w \not\in \llbracket \varphi \rrbracket \mbox{ and }\\ &\quad w' \in \llbracket \varphi \rrbracket\\
\end{cases}\\
I_{\Downarrow_P \varphi}\mbox{is the maximal subset of } I \mbox{ that is }\mathcal{A}-\mbox{coherent}\\
\end{array}
}
$
\end{defn}

As before, we can similarly define the lexicographic contraction on the desirability of $M$ by $\varphi$ (denoted by $M_{\Downarrow_D \varphi}$). 
These operations correspond to the contraction (or abandonment) of a belief/desire by the agent.

%

For each operation $\star$ defined before, we introduce a new modality $[\star\varphi]\psi$ in our language, meaning ``\textit{after the operation of} $\star$ \textit{by} $\varphi$, $\psi$ \textit{holds}''. which can be interpreted as
$$M,w\vDash [\star\varphi]\psi \qquad \mbox{iff} \qquad M_{\star\varphi},w\vDash \psi$$

An important result about the dynamified logic is that, if we consider some special kind of model, which includes the models we will use in Section~\ref{sec:reas} to reason about Agent Programming, it has the same expressibility as the static logic presented before \cite{souza2017dynamic}. In fact, the formulas $[!\varphi]\psi$, $[\Uparrow_\Box\varphi]\psi$ and $[\Downarrow_\Box\varphi]\psi$, with $\Box \in \{P, D\}$, are definable in the language $\mathcal{L}_{\leq_P,\leq_D}(\mathcal{A})$ by a set of reduction axioms. 

\section{Reasoning about BDI agents using DPL}
\label{sec:reas}

An interesting property of Preference Logic - the logic used as a foundation to construct $\mathcal{L}_{\leq_P,\leq_D}(\mathcal{A})$ - is that preference models can be encoded by means of some structures known as priority graphs \cite{liu2011reasoning}. Exploring this connection, we will show how we can use agent programs with stratified mental attitudes, e.g. beliefs annotated with their credence/plausibility level, to obtain a model for the agent's mental state.

In most BDI agent programming languages, an agent program is defined by means of a tuple $ag = {\langle K, B, D, I\rangle}$, where $K$, $B$ and $D$ are sets of (ranked) propositional formulas representing the agent's knowledge, beliefs and desires, respectively, and $I$ is a set of plans adopted by the agent. 

A ranked formula $\langle \varphi, n\rangle$ expresses that the agent has a certain degree of uncertainty in the information $\varphi$. As such, if the agent has $\langle \varphi, n\rangle$ in her belief base, it means that the agent believes that $\varphi$ is true with an uncertainty degree of $n$.

\begin{defn}
\label{def:agentprogram}
Let $P$ be a set of propositional variables and $\mathcal{A}$ a plan library. We call an agent program  over $\mathcal{A}$, a tuple of finite sets $ag = \langle K, B,D,I\rangle$ where:
\begin{itemize}
\item $K\subset \mathcal{L}_0$  is the knowledge base;
\item $B\subset \mathcal{L}_0\times \mathbb{N}$ is called the stratified belief base.
\item $D\subset \mathcal{L}_0\times \mathbb{N}$ is called a stratified goal base.
\item $I \subset \mathcal{A} $ is the (procedural) intention base.
\end{itemize}
When the plan library $\mathcal{A}$ is clear, we will often call the tuple $ag = \langle K, B,D,I\rangle$ an agent program.
\end{defn}


Given the definition above, we define the mental attitudes of an agent, i.e. what she knows, believes, etc. by means of the components of an agent program. 
Notice that our belief and desire bases are stratified, in the sense that the beliefs (similarly desires) of an agent are ranked according to their plausibility. Since some of these beliefs may be contradictory with each other, we must compute the maximal set of beliefs (desires) that is consistent - respecting the stratification of the base.

\begin{defn}\label{def:gammamax}
Let $\Gamma \subset \mathcal{L}_0\times \mathbb{N}$ be a finite set of pairs $\langle \varphi, i\rangle$ and let $\Gamma_i=\{\varphi ~|~ \langle \varphi, i\rangle \in \Gamma\}$. We define the maximal consistent subset of $\Gamma$, the set $\Gamma^{Max}\subset\mathcal{L}_0$, s.t.
\begin{itemize}
\item $\Gamma^{Max}\subseteq \bigcup \Gamma_i$ and if $\langle \varphi, i \rangle \in \Gamma$ and $\varphi \in \Gamma^{Max}$ then $\Gamma_i\subseteq \Gamma^{Max}$;
\item $ \forall \Gamma'\subseteq \bigcup_{i=1}^{\infty}\Gamma: (\exists \Gamma_i \subseteq \Gamma' \wedge \Gamma_i\not\subseteq \Gamma^{Max}\Rightarrow \Gamma'\vDash\bot \mbox{ or } \exists \Gamma_j \subseteq \Gamma^{Max} \wedge \Gamma_j\not\subseteq \Gamma' \mbox{ and } j<i)$
\end{itemize}
\end{defn}

With this in mind, we can provide interpretations of the notions of belief and desire by means of such bases.
%
%

\begin{defn}
\label{def:mental}
Let $ag = \langle K,B,D,I\rangle$ be an agent program and $\varphi \in \mathcal{L}_0$. We define what an agent believes, desires and intends as:
\begin{itemize}
\item $ag\vDash K\varphi$, iff $K\vDash\varphi$
\item $ag\vDash B\varphi$, iff $B^{Max}\vDash \varphi$
\item $ag\vDash G\varphi$, iff $D^{Max}\vDash \varphi$
\item $ag\vDash I\varphi$, iff $ag\vDash G\varphi$ and $\exists \alpha \in I$, s.t. $pos(\alpha)\vDash \varphi$
\end{itemize}
\end{defn}
%
%
%

While we placed no condition on agent programs, since in this work we adhere to Bratman's~\citeonline{bratman} notion of intention, our declarative mental attitudes must satisfy some constraints in order for the agent to be considered rational.

\begin{defn}
Let $ag = \langle K,B,D,I\rangle$ be an agent program. We say $ag$ is coherent iff all of the conditions below hold.
\begin{enumerate}
\item Knowledge consistency: $K\not\vDash \bot$
\item Belief-Knowledge consistency: $\varphi \in K$ iff $\langle \varphi ,0 \rangle \in B$
\item Desire-Knowledge consistency: $\varphi \in K$ iff $\langle \varphi ,0 \rangle \in D$
\item Intention-Desire consistency: for all $\alpha \in I$ there is a $\varphi$ in $D^{Max}$ s.t. $pos(\alpha) \vDash_{\mathcal{L}_0} \varphi$;
\item Pursuable plan: $\forall \alpha \in I: ag\vDash B(pre(\alpha))$;
\item Intention consistency: $\{pos(\alpha)~|~\alpha \in I\} \not\vDash\bot$;
\item Plans are relevant: $\forall \alpha \in I: ag\not\vDash B(pos(\alpha))$.
\end{enumerate}
\end{defn}

Liu~\citeonline{liu2011reasoning} shows that preference relations can be equivalently represented by means of syntactical constructs, known as priority graphs. A priority graph, however, is nothing more than a partial order over propositional formulas, much similar to the stratified bases we introduced here. As such, we can use the same reasoning to compute the plausibility and desirability orders of an agent model by means of belief and desire bases.

\begin{defn}\label{def:induc}
Let $\Gamma \subset \mathcal{L}_0\times \mathbb{N}$ be a stratified base, $W$ a set of possible worlds and $v: \mathcal{L}_0 \rightarrow W$ a valuation function. Considering $\Gamma_i = \{\varphi ~|~\langle \varphi , i\rangle \in \Gamma\}$ and $w\vDash X$ to stand for $\forall \varphi \in X
:(w\in v(\varphi))$, then we define the pre-order $\leq_\Gamma ~\subseteq ~W \times W$ s.t. 
\[
\begin{array} {lll}
w \leq_\Gamma w' \mbox{iff} & \forall i \in \mathbb{N}:& (w'\vDash \Gamma_i \Rightarrow w \vDash \Gamma_i) \vee\\
&&(\exists j<i \mbox{ s.t. } (w \vDash \Gamma_j \mbox{ and }w' \not\vDash \Gamma_j))
\end{array}
\]
\end{defn}

Using this construction, we are able to construct an agent model from an agent program.

\begin{defn}
Let $ag = {\langle K, B, D, I\rangle}$ be an agent program, we define the model induced by $ag$ as $M_{ag}= {\langle \llbracket K\rrbracket, \leq_B, \leq_D, I, v\rangle}$ where $\llbracket K \rrbracket \subset 2^P$ are all the propositional valuations that satisfy the set $K$, $\leq_B\subset \llbracket K \rrbracket \times \llbracket K \rrbracket$ and $\leq_D\subseteq \llbracket K \rrbracket \times \llbracket K \rrbracket$ are the preference relations induced by the bases $B$ and $D$, as described in Definition~\ref{def:induc}, and $w\in v(p)$ iff $p\in w$.
\end{defn}

Finally, since preference relations can always be encoded as priority graphs \cite{liu2011reasoning}, we can always compute agent programs describing mental models.

\begin{prop}
Let $M= \langle W, \leq_P \leq_D, I, v\rangle$ be an agent model, with $W\subseteq 2^P$, then there is an agent program $ag = \langle K,B,D,I\rangle$ s.t $M = M_{ag}$. More yet, $M$ is a coherent agent model iff $ag$ is a coherent agent program.
\end{prop}

From this result and the encodings of mental attitudes in both the logic and in agent programs, it is not difficult to see that the mental notions coincide.

\begin{corolary}
Let $ag = \langle K,B,D,I\rangle$ be a coherent agent program and $\varphi \in \mathcal{L}_0$ be a propositional formula, then $$\begin{array}{lll}
M_{ag} \vDash K(\varphi) &\mbox{ iff }& ag\vDash K\varphi\\
M_{ag} \vDash B(\varphi) &\mbox{ iff }& ag\vDash B\varphi\\
M_{ag} \vDash G(\varphi) &\mbox{ iff }& ag\vDash G\varphi\\
M_{ag} \vDash Int(\varphi) &\mbox{ iff }& ag\vDash I\varphi\\
M_{ag} \vDash I\alpha &\mbox{ iff }& \alpha \in I\\
\end{array}$$
\end{corolary}

We have two considerations to make about the codification presented here. First, regarding the complexity of reasoning about agent programs' attitudes, to compute an agents beliefs (or goals), it requires a linear number of propositional satisfiability checks on the depth of the base. Second, regarding the notion of mental attitudes encoded here, notice that  we adopted a notion of goal as a maximal set of consistent desires - consistent with other works in BDI programming. It is not difficult, however, to treat other notions, as that of Van Riemsdij et al~\citeonline{vanriemsdijk}, in our framework.

\section{Tractable Fragments of DPL}
\label{sec:trac}
We have seen so far that we can use the logic $\mathcal{L}_{\leq_P,\leq_D}(\mathcal{A})$ to reason about agent programs. The computational complexity of reasoning about agents, however, is far too great to be useful for real-world problems. 
In this section, we investigate some restrictions on the kinds of agent programs and agent models that guarantee that the reasoning problems in the resulting logic are tractable.

Since agent programs are defined over propositional formulas - and reasoning about propositional satisfiability is a well-known NP-complete problem - we define a restriction of agent programs for which reasoning will be proved to be tractable.

\begin{defn}
Let $\Gamma \subset \mathcal{L}_0\times \mathbb{N}$ be a stratified base, we say $\Gamma$ is conjunctive iff for all $\langle \varphi, i \rangle \in \Gamma$, $\varphi$ is a conjunction of literals, i.e. $\varphi = \bigwedge l_k$, with $l_k = p_k$ or $l_k =\neg p_k$.
\end{defn}

A conjunctive agent program is, thus, an agent program in which all of its bases are conjunctive.

\begin{defn}
Let $ag = {\langle K, B, D, I\rangle}$ be an agent program, we say $ag$ is a conjunctive agent program iff $K$ is a set of conjunctive formulas, $B$ and $D$ are conjunctive stratified bases and for any plan $\alpha \in I$, $pre(\alpha)$ and $pos(\alpha)$ are conjunctive formulas.
\end{defn}

First, we must show that we can compute the maximal consistent subsets, such as $B^{Max}$ and $D^{Max}$, in polynomial time. To do so, we provide the Algorithm $Max(\Gamma)$, depicted in Figure~\ref{fig:algMax}.

\begin{figure}[!ht]
\vspace{-.75\baselineskip}
\centering
\small
  \begin{lstlisting}[mathescape=true]
  $\ALG$ $Max(\Gamma)$
  $\INPUT$  $\mbox{a conjunctive stratified base } \Gamma$
  $\OUTPUT$ $\mbox{a set of literals }\Gamma^{Max} \mbox{ corresponding to the}$
         $\mbox{maximal consistent subset }  \mbox{ of the base } \Gamma$
    $\LIN{1}$   $\Gamma^{Max}$ := $\{\}$
    $\LIN{2}$   $n$ $:=$ $\mbox{maximal depth of } \Gamma$
    $\LIN{3}$   $\FOR$ $i$ $:=$ $1$  to $n$
    $\LIN{4}$      $\Gamma_i := \{l ~|~ \langle \varphi,i\rangle \in \Gamma \mbox{ and } l \mbox{ appears in } \varphi\}$
    $\LIN{5}$      $\IF$ $\neg l \in \Gamma_i$ $\AND$ $l \in \Gamma_i$ $\mbox{ for some } l$ $\THEN$
    $\LIN{6}$         $\mathbf{continue}$
    $\LIN{7}$      $\ELSE$
    $\LIN{8}$        $\IF$ $l \in \Gamma_i \AND \neg l \in \Gamma^{Max}\mbox{ for some } l$ $\THEN$
    $\LIN{9}$           $\mathbf{continue}$
    $\LIN{10}$        $\ELSE$
    $\LIN{11}$           $\Gamma^{Max}$ := $ \Gamma^{Max} \cup \Gamma_i$
    $\LIN{12}$  $\RETURN$ $\Gamma^{Max}$
  \end{lstlisting}
\caption{Algorithm for the maximal consistent subset of $\Gamma$.}
\label{fig:algMax}
\end{figure}

\begin{prop}
Let $\Gamma \subset \mathcal{L}_0\times \mathbb{N}$ be a conjunctive stratified base, then the algorithm $Max$ presented in Figure~\ref{fig:algMax} is correct, i.e. computes $\Gamma^{Max}$ in $O(n^3m^2)$ time, where $n$ is the size of $\,\Gamma$  and $m$ is the size of the biggest formula in $\Gamma$.
\end{prop}
As a consequence, we can always decide whether a conjunctive agent program knows (beliefs, desires or intends) a certain formula $\varphi$ in polynomial time.

\begin{corolary}\label{cor:k}
Let $ag = {\langle K, B, D, I\rangle}$ be a conjunctive agent program and $\varphi \in \mathcal{L}_0$ a propositional formula in disjunctive normal form. We can compute whether $ag\vDash K(\varphi)$ ($ag\vDash B(\varphi)$ or $ag\vDash D(\varphi)$) in polynomial time in the size of $K$ ($B$ or $D$) and $\varphi$.
\end{corolary}

Corollary~\ref{cor:k} guarantees that we can reason about the agent's mental state at any point in time in the program execution in polynomial time. The execution of an agent program, however, is usually determined by its reasoning cycle, i.e. the execution of certain mental changing operations that describe the agent's reasoning. These mental operations are usually described by means of changes in the agent's mental state. As such, to provide a truly tractable semantic framework to reason about agent programming, we must ensure that these mental changing operations can be computed in polynomial time. 

We now dedicate our attention to this problem. We aim to provide tractable operations on agent programs to compute the dynamic operations discussed in Section~\ref{sec:bdi}.

First, based on the work of Girard~\citeonline{girard2008modal} and of Liu~\citeonline{liu2011reasoning}, let's show how we can compute knowledge acquisition - interpreted here as a public announcement - using agent programs.

\begin{prop}
Let $ag = {\langle K, B, D, I\rangle}$ be an agent program, $\varphi\in \mathcal{L}_0$, and $ag' =  {\langle K\cup\{\varphi\}, B', D', I'\rangle}$, where $$\begin{array}{ll}
B' =& (B\cup\{\langle \varphi,0\rangle\})\\
D' =& (D\cup\{\langle \varphi,0\rangle\})\\
I' =& \{\alpha \in I ~|~ (B')^{Max} \vDash pre(\alpha) \mbox{ and } (B')^{Max} \not\vDash pos(\alpha)\\
&\mbox{ and } \qquad \quad\exists \varphi \in D': pos(\alpha)\vDash \varphi\}\end{array}$$ be the agent program resulting of agent $ag$ obtaining a knowledge $\varphi$. Then $M_{ag'} = {M_{ag}}_{!\varphi}$. We denote $ag'$ by $ag_{!\varphi}$.
\end{prop}

As a result of this encoding, we can compute knowledge acquisition/public announcement in polynomial time.

\begin{corolary}
Let $ag = {\langle K, B, D, I\rangle}$ be a conjunctive agent program and $\varphi,\psi \in \mathcal{L}_0$ conjunctive propositional formulas. We can compute whether $ag_{!\varphi}\vDash K(\psi)$ ($ag_{!\varphi}\vDash B(\psi)$ or $ag_{!\varphi}\vDash D(\psi)$) in polynomial time in the size of $K$ ($B$ or $D$), $\varphi$ and $\psi$.
\end{corolary}

As Radical Upgrade can also be computed by means of transformation on the agent programs, we can represent the mental operation of belief revision in our framework.

\begin{prop}
Let $ag = {\langle K, B, D, I\rangle}$ be a coherent agent program and $\varphi\in \mathcal{L}_0$, let yet $ag' =  {\langle K, B', D, I'\rangle}$, where
$$\begin{array}{ll}
B' =& \{\langle \psi, 0\rangle~|~ \psi \in K\}\cup\{\langle \psi,i+2\rangle~|~\langle \psi,i\rangle \in B \}\\
&\cup\{\langle \varphi,1\rangle\}\\
I' =& \{\alpha \in I ~|~ (B')^{Max} \vDash pre(\alpha) \mbox{ and } (B')^{Max} \not\vDash pos(\alpha)\}\end{array}$$
be the agent program resulting of agent $ag$ revising her beliefs by information $\varphi$. Then $M_{ag'} = {M_{ag}}_{\Uparrow_P\varphi}$. We denote $ag'$ by $ag_{\Uparrow_P \varphi}$.
\end{prop}

As a corollary, reasoning about the resulting mental state of the agent after belief revision is a tractable problem.

\begin{corolary}
Let $ag = {\langle K, B, D, I\rangle}$ be a conjunctive agent program and $\varphi,\psi \in \mathcal{L}_0$  conjunctive propositional formulas. We can compute whether $ag_{\Uparrow_P \varphi}\vDash K(\psi)$ ($ag_{\Uparrow_P \varphi}\vDash B(\psi)$ or $ag_{\Uparrow_P \varphi}\vDash D(\psi)$) in polynomial time in the size of $K$ ($B$ or $D$), $\varphi$ and $\psi$.
\end{corolary}

A similar result can be stated for the radical upgrade of the agents desires, instead of beliefs. This operation represents the adoption of a given goal.

To implement lexicographic contraction, we use the algorithm depicted in Figure~\ref{fig:algCont}. We represent by $\varphi[\psi|l]$ the substitution of literal $l$ appearing in $\varphi$ by the formula $\psi$.
\begin{figure}[!ht]
\vspace{-.75\baselineskip}
\centering
\small
  \begin{lstlisting}[mathescape=true]
  $\ALG$ $Cont(\Gamma, \varphi)$
  $\INPUT$  $\mbox{a conjunctive stratified base } \Gamma$
        $\mbox{a negated conjunctive formula}$ $\neg\varphi$
  $\OUTPUT$ $\Gamma_{\Downarrow \varphi}$ $\mbox{the lexicographic contraction of } \Gamma \mbox{ by } \varphi$
    $\LIN{1}$   $\Gamma_{\Downarrow \varphi}$ := $\{\}$
    $\LIN{1}$   $\FOREACH$ $\langle \psi, i\rangle \in \Gamma$
    $\LIN{2}$      $\psi'$ := $\psi$
    $\LIN{2}$      $\FOREACH 	\mbox{ propositional symbol } p \mbox{ in }\varphi$
    $\LIN{2}$         $\psi'$ := $\psi[\top|p][\top | \neg p]$ 
    $\LIN{3}$      $\Gamma_{\Downarrow \varphi}$ := $\Gamma_{\Downarrow \varphi} \cup \{\langle \psi', i\rangle\}$
    $\LIN{12}$  $\RETURN$ $\Gamma_{\Downarrow \varphi}$
  \end{lstlisting}
\caption{Algorithm for the contraction of a base $\Gamma$ by a formula $\varphi$.}
\label{fig:algCont}
\end{figure}

\begin{prop}
Let $ag = {\langle K, B, D, I\rangle}$ be a coherent agent program and $\varphi\in \mathcal{L}_0$, let yet $ag' =  {\langle K, B' , D, I'\rangle}$, where
$$\begin{array}{ll}
B' =& Cont(B, \varphi)\\
I' =& \{\alpha \in I ~|~ (B')^{Max} \vDash pre(\alpha) \mbox{ and } (B')^{Max} \not\vDash pos(\alpha)\}\end{array}$$
 be the agent program resulting of agent $ag$ contracting her beliefs by information $\varphi$. Then $M_{ag'} = {M_{ag}}_{\Downarrow_P\varphi}$, i.e. the algorithm $Cont(\Gamma, \varphi)$ depicted in Figure~\ref{fig:algCont} is correct. We denote $ag'$ by $ag_{\Downarrow_P \varphi}$.
\end{prop}

As before, we can reason about the changes in the mental state of the agent after the contraction of a belief ,or similarly the withdraw of a goal, in polynomial time to the size of the agent program and the formulas.

\begin{corolary}
Let $ag = {\langle K, B, D, I\rangle}$ be a conjunctive agent program and $\varphi\in \mathcal{L}_0$ a disjunctive propositional formula and $\psi \in \mathcal{L}_0$  conjunctive propositional formula, we can compute whether $ag_{\Downarrow_P \varphi}\vDash K(\psi)$ ($ag_{\Downarrow_P \varphi}\vDash B(\psi)$ or $ag_{\Downarrow_P \varphi}\vDash D(\psi)$) in polynomial time in the size of $K$ ($B$ or $D$), $\varphi$ and $\psi$.
\end{corolary}

With these results, we've provided a restriction of the logic which with which reasoning about agents' mental states is tractable and provided a way to translate from agent models to agent programs.

\section{Related Work}\label{sec:relwork}
From the Agent Programming perspective, the two most important works on modelling BDI mental attitudes are, in our opinion, the seminal work of Cohen and Levesque~\citeonline{Cohen90} and the work of Rao and Georgeff~\citeonline{RaoGeorgeff} describing the logic BDI-CTL. While their contribution to the area is undeniable, much criticism has been drawn to both approaches. Particularly, they are difficult to connect with agent programming languages, by the use of a possible-world model semantics.

Works as that of Wobcke~\citeonline{wobcke2004model} and of Hindriks and Van der Hoek~\citeonline{hindriks2008goal} propose ways to connect the semantics of a given programming language to some appropriate logic to reason about agent's mental attitudes. 
These logics, however, cannot represent the mental actions that characterize the practical reasoning process of the agent, i.e. the agent program execution. As such, it is not clear how to establish connection between the logic and agent programs.

Perhaps the work most related to ours in spirit is that of Hindriks and Meyer~\citeonline{hindriks2009toward}. They propose a dynamic logic for agents and show that this logic 
has an equivalent semantics based on the operational semantics of the programming language. The main difference between from our approach is that they choose to work in the framework of Situation Calculus and, as such, mental actions are only implicitly defined in their framework, while the inclusion of such actions in the language is exactly the main advantage advocated by us.  In some sense, our work can be seen as a generalisation of their work, since by employing Dynamic Preference Logic the equivalence they seek between operational semantics and declarative semantics can be automatically achieved by the results of Liu~\citeonline{liu2011reasoning}.


\section{Final Considerations}\label{sec:final}

Our work has investigated the use of a Dynamic Preference Logic to encode BDI mental attitudes and its connections to Agent Programming. We provided an expressive fragment of the logic for which reasoning about agents' mental states is tractable and how this can be computed by means of agent programs. With this, we believe we provided a roadmap to use Dynamic Preference Logic as a semantic framework to specify and also implement the formal semantics of BDI agent programming languages with declarative mental attitudes.

While we provide a fairly simple encoding of the mental attitudes in this work, the logic discussed here is expressive enough to encode different notions of desires, goals and intentions. For example, we can represent the semantics of goals as proposed by Van Riemsdijk et al \citeonline{vanriemsdijk} in our framework.



As a future work, we aim to implement a simple fragment of an agent programming language implementing declarative mental attitudes in this language by means of the codifications proposed in this work. We believe such an implementation can be used to understand the notions of mental attitudes imbued in the language. 

\bibliographystyle{IEEEtran}
\bibliography{ijcai17}

\end{document}